\documentclass{article}
\usepackage{moreverb}
\usepackage{siunitx, graphicx, amssymb}
\usepackage{amsmath, bm, array, booktabs, enumitem, makecell}
\usepackage{changepage, caption, subcaption, pdflscape,setspace}
\usepackage{lscape}
\begin{document}

\begin{titlepage}

\begin{center}

{\LARGE \bf Practical Guidance on Modeling Choices for the Virtual Twins Method} \\
\ \\
{\large Chuyu Deng$^{a, \dagger}$, David M. Vock$^{a}$, Dana M. Carroll$^{b}$, Jeffrey A. Boatman$^{a}$, Dorothy K. Hatsukami$^{c, d}$, Ning Leng$^{e}$, and  Joseph S. Koopmeiners$^{a}$ }\\
\ \\
$^{a}$ Division of Biostatistics, School of Public Health, University of Minnesota, Minneapolis, MN 55455, USA\\
$^{b}$ Division of Environmental Health Sciences, School of Public Health, University of Minnesota, Minneapolis, MN 55455, USA\\
$^{c}$ Masonic Cancer Center, University of Minnesota, Minneapolis, MN 55455, USA \\
$^{d}$ Department of Psychiatry and Behavioral Sciences, University of Minnesota, Minneapolis, MN 55455, USA \\
$^{e}$ Genentech, San Francisco, CA 94080, USA \\
\end{center}

\begin{abstract}
Individuals can vary drastically in their response to the same treatment, and this heterogeneity has driven the push for more personalized medicine. Accurate and interpretable methods to identify subgroups that respond to the treatment differently from the population average are necessary to achieving this goal. 
The Virtual Twins (VT) method by Foster et al. \cite{Foster} is a highly cited and implemented method for subgroup identification because of its intuitive framework. 
However, since its initial publication, many researchers still rely heavily on the authors' initial modeling suggestions without examining newer and more powerful alternatives. This leaves much of the potential of the method untapped. 
We comprehensively evaluate the performance of VT with different combinations of methods in each of its component steps, under a collection of linear and nonlinear problem settings. 
Our simulations show that the method choice for step 1 of VT is highly influential in the overall accuracy of the method, and Superlearner is a promising choice. We illustrate our findings by using VT to identify subgroups with heterogeneous treatment effects in a randomized, double-blind nicotine reduction trial. 
\end{abstract}

$^{\dagger}${\bf Corresponding author}: Chuyu Deng, Division of Biostatistics, University of Minnesota, Minneapolis, MN, 55455, email: dengx361@umn.edu \\
{\bf Acknowledgements}: This work was partially supported by the National Cancer Institute (Award Numbers R01CA214825 and R01CA225190), the National Institute on Drug Abuse (Award Numbers R01DA046320, R03DA041870, and U54-DA031659) and National Center for Advancing Translational Science (Award Number UL1TR002494). The content is solely the responsibility of the authors and does not necessarily represent the official views of the National Institutes of Health or Food and Drug Administration.


\end{titlepage}

\newpage
\setcounter{page}{3}

\section{Introduction}

In healthcare, the traditional treatment (intervention/action/etc.) paradigm has historically been a one-size-fits-all framework, where all patients with a certain disease are given the same treatment regardless of their baseline attributes. Yet, we know that individuals can vary drastically in their response to treatments: some people might benefit substantially from a treatment, some may only benefit marginally or not at all, and others could be harmed. This heterogeneity in treatment effects has led to a push for replacing the one-size-fits-all framework with personalized medicine, where individuals are given treatments personalized to their unique characteristics, in which case everyone, in theory, will be receiving the best treatment for their individual patient characteristics. 

In order for medicine to move to this personalized healthcare framework, we must first understand how different individuals will respond to the same treatment, or treatment effect heterogeneity. Varadhan et al.\@ defined treatment effect heterogeneity (TEH) as nonrandom explainable variability in the direction and magnitude of individual treatment effects \cite{Varadhan}. TEH is commonly evaluated using either confirmatory (hypothesis testing) or exploratory (hypothesis generating) analyses. In confirmatory analyses, the goal is to rigorously test pre-specified hypotheses. By comparison, analyses that lack a pre-specified hypothesis are usually deemed exploratory, where the goal is to generate new hypotheses for future testing. However, the line between this distinction is often blurred and evidence from exploratory analyses are sometimes used as conclusive evidence. 

Even when confirmatory analyses are done correctly, adhering to the “best practices for subgroup analysis”, there are still problems as outlined by Lipkovich et al.\@ \cite{Ilya}: Under the conventional guidelines, all subgroups must be pre-specified to prevent data dredging, but this forces scientists to choose only a few hypotheses to test, and could severely limit the scope of the scientific inquiry. Another common guideline is that no testing in a subgroup should be done unless the corresponding interaction test is significant. This assumes that modeling of the main effects and interactions is being performed in a one-predictor-at-a-time fashion and fails to consider other more efficient methods of modeling. Finally, it’s often suggested that no testing in subgroups should be performed unless the overall effect is significant, which seems contrary to the idea of personalized medicine where the goal is to find subgroups where the treatment effect is significant, even if it's not in other subgroups. Lipkovich et al.\@ suggests that instead of an unnecessarily restrictive guideline-driven approach, we should be using a pre-specified and principled, but data-driven approach \cite{Ilya}. 

Currently, there are several existing methods in the literature that perform data-driven subgroup identification, including tree-based methods, hybrid methods, and ensemble methods. 
Tree-based methods range from older methods such as classification and regression trees (CART) \cite{Breiman}, to newer methods such as honest causal trees \cite{Athey}. In honest causal trees, the “causal effects” refer to the individual treatment effects obtained by taking the difference between the outcome we observe under one treatment, and the counterfactual outcome that would have been observed had the individual received the other treatment. The “honesty” condition refers to the fact that the data used to predict which covariates dictate the splits in the trees are not used to estimate the covariates in the tree. The training set is split into two parts, and one is used to construct the tree, while the other is used to estimate the treatment effects within the leaves of the tree. Although there is a loss of precision due to the sample splitting in the training set, this method allows the elimination of bias that could offset some of the loss. 
Hybrid methods are methods that combine various existing models to create new models with better performance. FindIt \cite{Imai} adapts support vector machine (SVM) by placing 2 separate least absolute shrinkage and selection operator (LASSO) sparsity constraints over the pre-treatment covariates: one for the predictive covariates (covariates that are predictive of the treatment effect), and another for prognostic covariates (covariates that are predictive of the disease/outcome prognosis). This approach accounts for the fact that the predictive effects are inherently weaker and need to be treated differently from the prognostic effects. Essentially, 2 separate LASSO penalties are added to an SVM loss function to perform simultaneous variable selection, which allows for a simpler interpretation in the outcome. 
The Simultaneous Threshold Interaction Modeling Algorithm (STIMA) \cite{Dusseldorp} combines additive models with tree-based regression models to both capture linear main effects of continuous predictors and higher order interaction effects. A linear model is fit at the trunk of the tree, and a pruned tree is made into the form of an additive model by using indicators for all the splits that lead to every node. This creates a hybrid linear/tree-based model that is interpretable and more parsimonious than a linear model. Simulation results for STIMA were comparable to existing methods such as multivariate adaptive regression splines (MARS) \cite{Friedman2} and Generalized, Unbiased, Interaction Detection and Estimation (GUIDE) \cite{Loh}. 
Finally, there are ensemble methods such as Random Forests \cite{Breiman2} and Superlearner \cite{Grimmer} that combine the results from many simpler machine learning methods to improve predictive accuracy. 

Generally, tree-based and hybrid methods like those previously mentioned give interpretable results that could be used in practice by clinicians. However, these results are often not as accurate or robust as results from black-box ensemble methods such as Superlearner. To combine the predictive accuracy of more complicated ensemble methods and the interpretability of simpler methods, the Virtual Twins framework (VT) was developed by Foster et al.\@\cite{Foster}. 
VT is a flexible framework that separates estimation of the individual treatment effects from estimation of the decision rule model. This allows the use of a more accurate, albeit more complicated, model for the estimation of the treatment effects, while preserving the interpretability of a simpler method for delineating the final subgroups. But, while the flexible specification of models in VT can give both an interpretable and accurate final model, this also burdens the user with making appropriate modeling choices. As our simulations prove, this modeling choice selection is crucial in the performance of VT. 

Since its publication, VT has been highly cited and applied in a variety of settings because of its intuitive framework. In the original Foster et al.\@ paper, the authors demonstrated the promise of VT with a logistic model and random forest model. While Foster et al.\@ noted that many other variations could be implemented, many researchers rely heavily on these initial suggestions without examining alternatives.
To date, no one has detailed a comprehensive analysis of the different modeling choices that can affect the performance of VT in variety of real-world settings, or the performance of newer predictive methods in the VT framework. 
This leaves many researchers hesitant to deviate from methods proposed by Foster et al.\@, and leaves much of the potential of VT untapped.
To address this gap in knowledge, we conduct a simulation study to evaluate the performance of VT with different combinations of methods in each of its component steps, under a collection of linear and nonlinear problem settings in a comprehensive fashion.

\section{Methods}
We first provide a brief overview of the VT algorithm, first published by Foster et al.\@ \cite{Foster}. A complete description of the method can be found in the original manuscript. We then detail the modeling choices faced by researchers when using VT. 

\subsection{Notation}
The data for each individual consist of a continuous outcome $Y_i$, a binary treatment indicator $T_i=j$ where $j=1$ denotes treatment and $j=0$ denotes control, and covariates $1...p$ denoted by the matrix $\boldsymbol{X_i}=X_{1i},...,X_{pi}$. 
We define $\bm\tilde{Y}_{1i}=E(Y_i|T_i=1, \boldsymbol{X_i})$ and $\bm\tilde{Y}_{0i}=E(Y_i|T_i=0, \boldsymbol{X_i})$ for each individual i. 
The goal of elucidating TEH is to find subgroups $S(\boldsymbol{X})$ of individuals defined by a rule that subsets the overall population based on the vector $\boldsymbol{X}$. Virtual Twins achieves this goal by looking for subgroups in the individual treatment effects, or $Z_i$, of a population.  
Although we will focus on the situation where we have a continuous outcome and binary treatment assignment, this framework can be easily extended to other types of endpoints as well.

\subsection{Virtual Twins} \label{VT}

\captionsetup[sub]{font+=large} 
\label{vtsteps}
\begin{figure}[bt]
\centering
\subcaptionbox{Step 1: building individual response surfaces}{\includegraphics[width=0.45\textwidth]{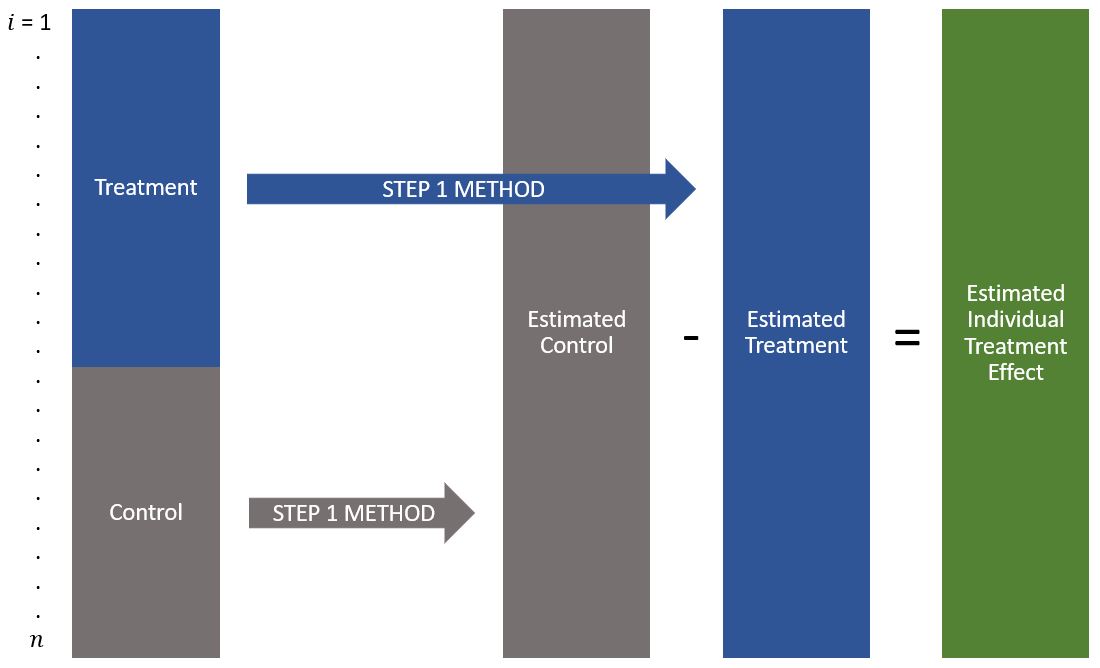}}\label{vt1}
~
\subcaptionbox{Step 2: identifying subgroups}{\includegraphics[width=0.45\textwidth]{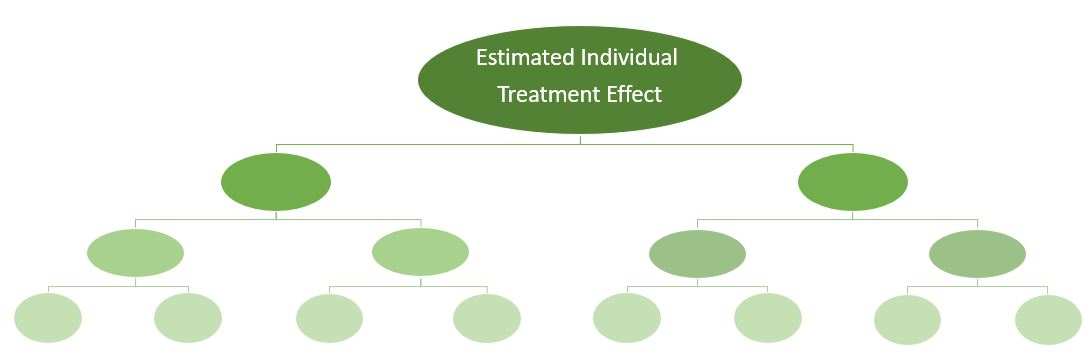}}
\caption{Overview of the 2 main steps in the Virtual Twins method}
\end{figure}

The Virtual Twins framework is a two-step estimation approach that has its roots in the potential outcomes models of causal inference. As outlined in Figure \ref{vtsteps}, the process consists of first building the individual response surfaces, and then identifying heterogeneous subgroups based on the estimated response surfaces. 
In the first step, we build the conditional expectation functions $f(.)$ for predicting outcomes in each treatment arm using only the covariate information for individuals in each respective arm. With this function, we obtain the estimates
$\bm\tilde{Y}_{0i}=f(\boldsymbol{X}_i, T=0)$ and $\bm\tilde{Y}_{1i}=f(\boldsymbol{X}_i, T=1)$ for each subject $i=1...n$. We can then calculate an estimate for the individual treatment effect as $\bm\tilde{Z}_i=\bm\tilde{Y}_{1i}-\bm\tilde{Y}_{0i}$, which is the difference in expected outcomes if an individual was in the treatment arm vs the control arm. In the second step of VT, we fit a tree structure with $\bm\tilde{Z}$ as the response variable and covariates $X$. The tree is used to partition subjects into heterogeneous groups based on their estimated treatment effects $\bm\tilde{Z}_i$. Thus, the resulting subgroups are defined by the splits in the tree. 

\begin{figure} [bt]
\centering
\includegraphics[width=0.7\textwidth]{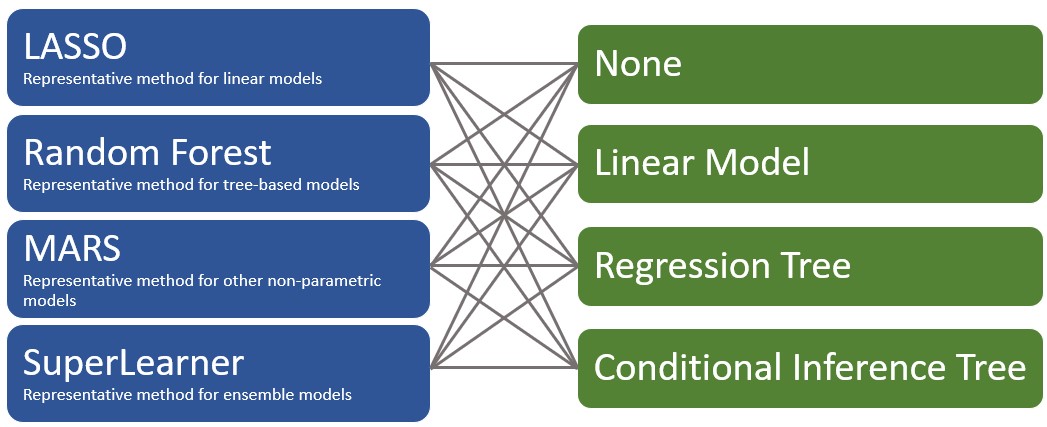}
\caption{All the combinations of Step 1 and Step 2 methods we will evaluate in our simulations.}
\label{combos}
\end{figure}

A wide variety of models can be applied in both steps of the VT framework. In our simulations, we will consider the different methods for performing step 1 and step 2 shown in Figure \ref{combos} to evaluate the effect on VT's performance. We evaluate all possible combinations between the four step 1 and 2 models to understand the effect of each combination on the performance of VT. 

\subsection{Step 1 of Virtual Twins} 

This section describes the methods that were compared for step 1 of VT, where we are building the individual response surfaces. We will also clarify the modeling choices that were made within each particular method. In step 1 of VT, we explore a variety of different models, including:

\paragraph{LASSO} 
LASSO \cite{Tibshirani} is a linear regression method that performs variable selection and regularization simultaneously by penalizing the absolute size of the regression coefficients. When LASSO is used in the first step, we estimate the underlying regression functions $f(.)$ using LASSO regression models. We used the \texttt{cv.glmnet} function from the \texttt{glmnet} package in R  with default settings \cite{glmnet}, except that \texttt{lambda.1se} was used as the cross validated tuning parameter to encourage model sparsity. 

\paragraph{Random Forest}
Random Forest \cite{Breiman2} is an ensemble method that averages the results of many relatively uncorrelated decision trees to output a robust solution. When Random Forest is used in the first step, we estimate the underlying regression functions $f(.)$ using Random Forest models. The \texttt{tune} function from the \texttt{randomForestSRC} package is used in R \cite{rfsrc}, so the forest is automatically tuned for the number of variables randomly sampled as candidates at each split (\texttt{mtry}), and the minimum size of terminal nodes (\texttt{nodesize}). All other default settings were kept. 

\paragraph{MARS} 
MARS \cite{Friedman3} is a non-parametric regression method that creates a model using spline basis functions with automatic feature selection. It can be seen as a generalization of CART methods that is better at handling numeric variables. When MARS is used in the first step, we estimate the underlying regression functions $f(.)$ using MARS models. The \texttt{earth} package in R was used with all default settings \cite{earth}. 

\paragraph{Superlearner} 
Superlearner \cite{Van} is an ensemble method that uses cross-validation to determine the optimal combination of several candidates models to form a single predictive model.When Superlearner is used in the first step, we estimate the underlying regression functions $f(.)$ using Superlearner models. The \texttt{Superlearner} package in R \cite{sl} was used with all default settings, and with the following wrappers: \texttt{SL.glmnet}, \texttt{SL.randomForest}, and \texttt{SL.earth}. Note that the \texttt{SL.earth} wrapper was slightly modified to match the default settings of the \texttt{earth} package. These specific wrappers were used so that the Superlearner models would encompass all the other candidate models for step 1 of VT.

\subsection{Step 2 of Virtual Twins} 

This section describes the methods that were compared for step 2 of VT, where we are identifying subgroups for TEH. We will also clarify the modeling choices that were made within each particular method. In step 2 of VT, we evaluated several cases: 

\paragraph{None}
No step 2 method was used. The estimated individual treatment effects $\bm\tilde{Z}_i$ were used without any subgroup assignments for the individual treatment benefit classification accuracy, and for the individual treatment benefit MSE. When prediction accuracy is of paramount importance, not using a second step would be expected to yield the best results. This also allows us to quantify the cost in accuracy for other models that favor interpretability. 

\paragraph{Linear Model}
A linear model was used. Only the top predictors from a cross-validated LASSO are used in the linear model as main effects, and the number of top predictors is set to be the same as the number of predictors selected by the regression tree. Although linear models do not allow for a natural way to delineate multiple subgroups, they are one of the most commonly used and interpretable models available. This method also serves as a benchmark for the evaluation of the tree based step 2 methods. Only the top predictors from a cross-validated LASSO (again using the \texttt{glmnet} package \cite{glmnet}) are used in the linear model, and the number of top predictors is set to be the same as the number of predictors selected by the regression tree. 

\paragraph{Regression Tree}
A regression tree was used. Regression trees are a natural choice for step 2 of VT since it allows us to easily create interpretable subgroups from the individual treatment effects. This is the primary method mentioned in the original VT paper for step 2 of VT. The \texttt{caret} package in R was used \cite{caret} with default settings for the \texttt{rpart2} model. The maximum depth of the tree was tuned using a 10 fold cross-validation, repeated 3 times. 

\paragraph{Conditional Inference Tree}
A conditional inference tree \cite{Hothorn} was used. Conditional inference trees use a significant test procedure for variable selection to avoid selecting variables that have many possible splits or many missing values. Conditional inference trees are a variation on regression trees that should have better performance for variable selection. The \texttt{caret} package in R was used \cite{caret} with default settings for the \texttt{ctree2} model. The maximum depth of the tree was tuned using a 10 fold cross-validation, repeated 3 times. 

\subsubsection{Parameter Tuning} 
To control the step 2 model's behavior in the absence of TEH we propose a permutation-based method to identity tuning parameters for the step 2 models so that a covariate is included in the model with probability $\le \alpha$.
This approach supports LASSO, regression trees, and conditional inference trees.

To do so, we permute the treatment indicators ($T_i$) in our data and refit the selected step 1 model on this permuted data. 
We then identify the smallest possible penalty parameter for the stage 2 model (\texttt{lambda} for \texttt{glmnet} from \texttt{glmnet} package, \texttt{cp} for \texttt{rpart} from the \texttt{rpart} package, and \texttt{mincriterion} with \texttt{testtype = "Teststatistic"} for \texttt{ctree} from the \texttt{party} package) such that the model contains no covariates (e.g. an intercept-only LASSO).
We repeat this process $M$ times and take the $1-\alpha$ quantile of all returned penalty parameters to use in the step 2 model on the real data.

\section{Simulations}
\label{sims}
We completed a thorough simulation study to evaluate the performance of the Virtual Twins framework under different data generation schemes and modeling choices. 

\subsection{Settings}
\label{settings}
We evaluated the performance of VT under both linear and nonlinear data generation scenarios. 
For both cases, 3 different data generation models were considered: 
1.	Regular case - where the covariates are linearly/nonlinearly related to the outcome. 
2.	Correlated covariates case - where some of the covariates are correlated to each other, but only one is related to the outcome.
3.	Selection bias case - where the covariate distribution of the training sample and testing samples are different. 
For each data generation model we considered a case where 1. there is TEH and 2. there is no TEH.
These problem settings are designed to mimic an ideal dataset, and real data problems that can frequently occur in datasets, such as correlated features in high dimensions, and selection bias. 
We are assuming a continuous outcome measure, with a binary treatment assignment and both continuous and categorical covariates. Specifics on the data generation can be found in Appendix \ref{app_DG}. 
In addition to the data generation scheme, the sample size was also varied at 3 different levels: N=1000, N=600 and N=200. 
Overall, there were a total of 9 different problem settings for each of the linear and nonlinear cases.

\subsection{Summaries of Performance}
To measure the performance of the different VT method combinations under the various problem settings, we used 3 metrics.

\paragraph{Classification Accuracy}
For the individual treatment benefit classification, the estimated optimal treatment group for each subject at the end of the second step of VT is compared to their true optimal treatment group. For each subject $i$, $argmax_{T\in\{0,1\}}\bm\tilde{Y}_{Ti}$ is taken to be the estimated optimal treatment, and $argmax_{T\in\{0,1\}} Y_{Ti}$ is taken to be the true optimal treatment. The correct classification rate for each simulation is defined as:
$\frac{1}{n}\sum_{i=1...n} I(argmax_{T\in\{0,1\}}\bm\tilde{Y}_{Ti} = argmax_{T\in\{0,1\}} Y_{Ti})$. 
The correct classification rate averaged over 1000 simulations are reported below. 

\paragraph{Individual treatment effect MSE}
In addition to the optimal treatment group classification, we are also interested in how accurate the individual treatment effect estimates, $\bm\tilde{Z}_i$, are compared to the true individual treatment effects, $Z_i$. The individual treatment effect MSE for each simulation is defined as: 
$\frac{1}{n}\sum_{i=1...n} (\bm\tilde{Z}_i-Z_i)^2$ and the MSEs reported in the tables below are averaged over 1000 simulations. 

\paragraph{Proportion of variables correctly selected by each model}
The final metric we are interested in is if the step 2 models are able to isolate the variables responsible for the TEH. To calculate the proportion of variables correctly selected by each model, the covariates selected into the cross-validated tree models and linear models of step 2 are compared to the true predictive covariates. Since the step 2 method of ``None" does not use any covariates, this was omitted. The pooled results for the variables selected from 1000 simulations are reported.

\subsection{Results}
We divide the simulation results into separate sections for the linear and nonlinear data generation cases. We present the simulation results for the linear case in Section \ref{lin.res} and the results for the nonlinear case in Section \ref{nonlin.res}. Additional linear and nonlinear simulation results with a sample size of N=80 can be found in Appendix \ref{app_80}.

\subsubsection{Linear Case Results}
\label{lin.res}
\begin{table} [!ht]
\centering
    \begin{subfigure}{0.6\textwidth}
        \includegraphics[width=\textwidth]{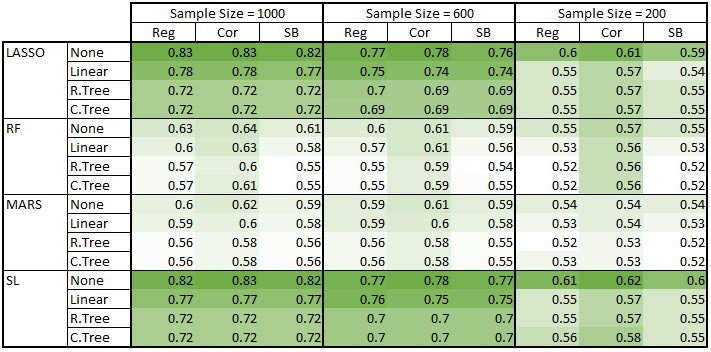}
        \caption{Individual treatment benefit classification accuracy}
        \label{linearwg}
    \end{subfigure}

    \begin{subfigure}{0.6\textwidth}
        \includegraphics[width=\textwidth]{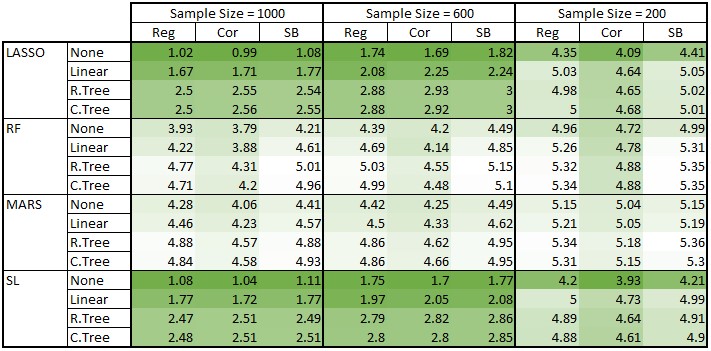}
        \caption{Individual treatment effect MSE}
        \label{linearmse}
    \end{subfigure}
    
    \begin{subfigure}{0.6\textwidth}
        \includegraphics[width=\textwidth]{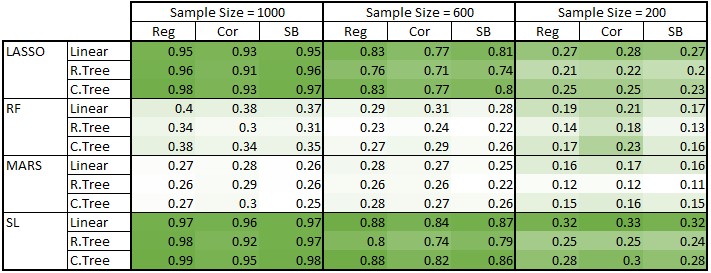}
        \caption{Proportion of variables correctly selected by each model}
        \label{linearvar}
    \end{subfigure}
\caption{Results of linear case simulations under varying scenarios for 3 metrics. The rows indicate the methods used for step 1 and 2 of VT and the columns denote the different sample sizes and data settings. The labels of ``Reg", ``Cor" and ``SB" indicate the data generation of the regular case, correlated case and selection bias case described in section \ref{settings}. The labels of ``SL", ``R.Tree" and ``C.Tree" refer respectively to the Superlearner, regression trees and conditional inference trees used in VT.}
\end{table}

In the linear case, the performance of VT is highly influenced by the method used in the first step of VT across all 3 metrics. 
The individual treatment benefit accuracy rates for all the problem settings and VT method combinations are displayed in Table \ref{linearwg}. With a sample size of 1000 or 600, using the LASSO or Superlearner for step 1 of VT gave the most accurate classification of $\sim$80\% when no step 2 was used. When a linear model is used for step 2 in conjunction with LASSO or Superlearner as the step 1 method, the classification accuracy drops to $\sim$75\%. When a tree based method is used for step 2 in conjunction with LASSO or Superlearner as the step 1 method, the accuracy drops again to $\sim$70\%. Using Random Forest or MARS for step 1 gave classification accuracy rates of $\sim$60\%.
With a sample size of 200, using the LASSO or Superlearner for step 1 and using no step 2 model yield a classification accuracy of $\sim$60\%. Changing the step 1 or step 2 method in this case to any other method leads to a classification accuracy of $\sim$55\%. 
The different problem settings did not have a large impact on the classification accuracy. 

The individual treatment MSEs are displayed in Table \ref{linearmse}. The pattern is consistent with the results for classification accuracy, where using LASSO or Superlearner for Step 1 had the lowest MSEs. Using no step 2 method was again best, with a small penalty for using a linear model, and a bigger penalty for using a tree-based model. 

The proportion of variables correctly selected by the models are displayed in Table \ref{linearvar}. Here, we see a drastic difference between the step 1 methods of LASSO and Superlearner versus Random Forest or MARS:
With a sample size of 1000 or 600, the variables picked out by both the LASSO and Superlearner are $\sim$90-99\% true predictors for all the problem settings, as opposed to the $\sim$20-40\% when Random Forest or MARS are used. Using conditional inference trees for step 2 gave marginally better results than the linear model and regression tree in the 1000 sample size case. However with a sample size of 600, the conditional inference trees and linear models outperform regression trees. 
With a sample size of 200, using the Superlearner with a linear model in step 2 yielded the best results, but all the method variations gave results of $\sim$15-33\%.

\subsubsection{Nonlinear Case Results}
\label{nonlin.res}
\begin{table} [!htbp]
\centering
    \begin{subfigure}{0.6\textwidth}
        \includegraphics[width=\textwidth]{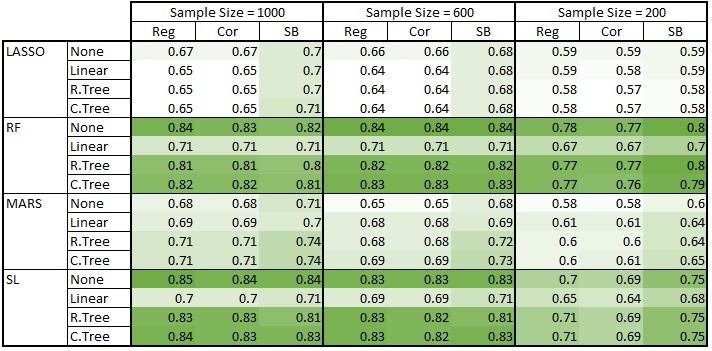}
        \caption{Individual treatment benefit classification accuracy in nonlinear case}
        \label{nonlinearwg}
    \end{subfigure}

    \begin{subfigure}{0.6\textwidth}
        \includegraphics[width=\textwidth]{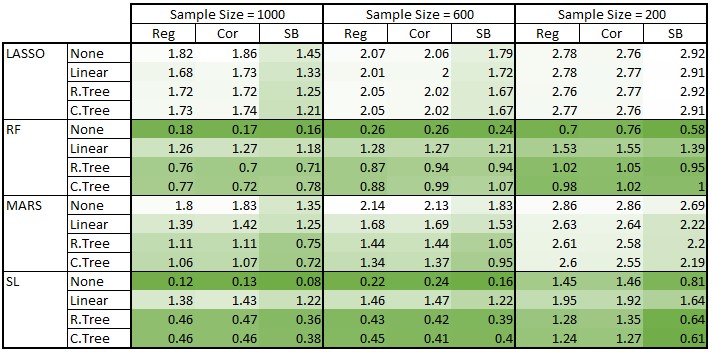}
        \caption{Individual treatment benefit MSE in nonlinear case}
        \label{nonlinearmse}
    \end{subfigure}
    
    \begin{subfigure}{0.6\textwidth}
        \includegraphics[width=\textwidth]{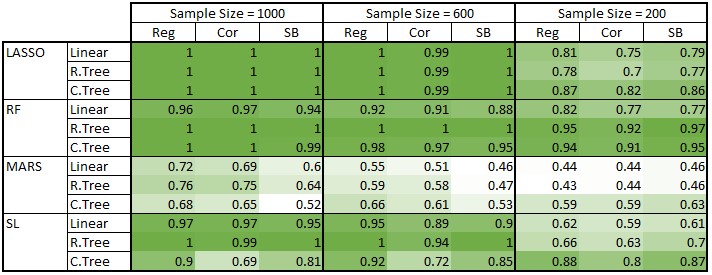}
        \caption{Proportion of variables correctly selected by each model in nonlinear case}
        \label{nonlinearvar}
    \end{subfigure}
\caption{Results of nonlinear case simulations under varying scenarios for 3 metrics. The rows indicate the methods used for step 1 and 2 of VT and the columns denote the different sample sizes and data settings. The labels of ``Reg", ``Cor" and ``SB" indicate the data generation of the regular case, correlated case and selection bias case described in section \ref{settings}. The labels of ``SL", ``R.Tree" and ``C.Tree" refer respectively to the Superlearner, regression trees and conditional inference trees used in VT.}
\end{table}

In the nonlinear case, we can again see that the performance of VT is highly influenced by the method used in step 1 of VT. The individual treatment benefit accuracy rates for all scenarios and VT method combinations are displayed in Table \ref{nonlinearwg}. 
With a sample size of 1000 and 600, using the Random Forest or Superlearner for step 1 of VT gave the most accurate classification of $\sim$85\% for every step 2 method except for the linear model, where the classification accuracy drops to $\sim$70\%. All other method combinations performed similarly at $\sim$65-75\% classification accuracy. 
At a sample size of 200, using Random Forest for step 1 of VT still gave results of $\sim$80\% accuracy for every step 2 method except for the linear model, where the classification accuracy drops to $\sim$70\%. Superlearner performed well, with an accuracy of $\sim$70\%. Using LASSO or MARS for step 1 gave similar performances of $\sim$60\% classification accuracy. 

The individual treatment MSEs are displayed in Table \ref{nonlinearmse}. Here again, the patterns of the results are consistent with the results from the classification accuracy, where using Random Forest or Superlearner for step 1 of VT gave the best results throughout the different sample sizes. The choice in step 2 method was also important here since the linear model performed drastically worse than the tree-based methods. 

The proportion of variables correctly selected by the models are displayed in Table \ref{nonlinearvar}. 
For sample sizes of 1000 and 600, when LASSO, Random Forest or Superlearner are used as the step 1 method, $\sim$95-100\% of variables selected were true predictors, with the exception of the correlated covariates scenario using conditional inference trees as the step 2 method. Here, the proportion of variables correctly selected is only $\sim$70\%. 
When MARS is used as the step 1 method, only $\sim$50-75\% of the variables picked were true predictors for all three step 2 methods.
At a sample size of 200, using Random Forest as the step 1 method and a tree-based method for step 2 had the best results at a true predictor detection rate of $\sim$95\%. When a linear model was used as the step 2 method after Random Forest, $\sim$80\% of selected variables were true predictors. In the case of LASSO as the step 1 method, $\sim$75-85\% of variables selected were correct. Superlearner only performed well as a step 1 method when paired with conditional inference trees, giving a correct selection rate of $\sim$85\%. For all other method combinations, only $\sim$45-65\% of selected variables were true predictors.


\section{Data Application} 
\label{application}
In clinical trials, smokers randomized to receive very low nicotine content (VLNC) cigarettes versus normal nicotine content cigarettes (control) exhibit, on average, significantly reduced cigarette use, dependence, and biomarkers of harmful constituent exposure compared to the control \cite{hatsukami2010reduced, benowitz2012smoking, benowitz2015effect, hatsukami2013reduced}. In our data application, we use VT to evaluate the existence of TEH in a randomized trial of VLNC cigarettes.  

\begin{table}[!htbp]
\caption{Demographic and baseline characteristics of trial participants.}
\footnotesize
\centering
\def\arraystretch{0.9}
\begin{tabular}{lcc}
\toprule
\textbf{Variable} & \multicolumn{2}{c}{\textbf{Median[IQR] or Count(\%)}} \\
\midrule
& \textbf{Control} & \textbf{VLNC} \\ 
  n & 198 & 340 \\ 
  Age & 47 [35, 55] & 49 [36, 57] \\ 
  Gender = Male & 108 (54.5) & 182 (53.5) \\ 
  Baseline Weight (kg) & 81 [68, 92] & 85 [71, 100] \\ 
  Race &  &  \\ 
  \qquad    White & 121 (61.1) & 215 (63.2) \\ 
  \qquad    Black & 61 (30.8) & 103 (30.3) \\ 
  \qquad    Other & 16 (8.1) & 22 (6.5) \\ 
  Education &  &  \\ 
  \qquad    HS or less & 18 (9.1) & 31 (9.1) \\ 
  \qquad    HS Grad & 65 (32.8) & 105 (30.9) \\ 
  \qquad    College or more & 115 (58.1) & 204 (60.0) \\ 
  Use of Menthol Cigarettes = Non-menthol & 113 (57.1) & 182 (53.5) \\ 
  Baseline Total Cigarettes/Day & 16 [10, 21] & 16 [11, 22] \\ 
  Baseline Expired Carbon Monoxide (ppm) & 18 [12, 25] & 17 [13, 24] \\ 
  Baseline Total Nicotine Equivalents (nmol/mL) & 64 [40, 96] & 64 [40, 93] \\ 
  Biomarker for NNK Exposure (pmol/mg) & 2 [1, 2] & 1 [1, 2] \\ 
  Biomarker for Acrylonitrile Exposure (nmol/mg) & 1 [0, 1] & 1 [0, 1] \\ 
  Phenanthrene tetraol (pmol/mg) & 2 [1, 4] & 2 [1, 4] \\ 
  PGEM (pmol/mL) & 43 [28, 84] & 50 [33, 75] \\ 
  8-isoPGF2a (pmol/mL) & 1 [1, 2] & 1 [1, 2] \\ 
  FTND Score without CPD & 4 [3, 5] & 4 [3, 6] \\ 
  CESD Score & 5 [2, 9] & 5 [2, 8] \\ 
  SMAST Score & 3 [2, 4] & 3 [1, 4] \\  
  DAST Score & 1 [0, 2] & 2 [0, 2] \\ 
  Perceived Stress Scale & 4 [2, 7] & 4 [2, 6] \\ 
  WISDM Total Score & 39 [30, 49] & 38 [30, 47] \\ 
  \quad WISDM Primary Dependence Motives & 4 [3, 5] & 4 [3, 5] \\ 
  \qquad WISDM Automaticity & 4 [2, 5] & 4 [2, 5] \\ 
  \qquad WISDM Loss of Control & 4 [2, 5] & 4 [2, 5] \\ 
  \qquad WISDM Craving & 4 [3, 6] & 4 [3, 6] \\ 
  \qquad WISDM Tolerance & 5 [4, 6] & 5 [4, 6] \\ 
  \quad WISDM Secondary Dependence Motives & 3 [2, 4] & 3 [2, 4] \\ 
  \qquad WISDM Affiliative Attachment & 2 [1, 3] & 2 [1, 3] \\ 
  \qquad WISDM Cognitive Enhancement & 3 [1, 4] & 2 [1, 4] \\ 
  \qquad WISDM Cue Exposure & 4 [3, 5] & 4 [3, 5] \\ 
  \qquad WISDM Social Goads & 4 [2, 6] & 3 [2, 5] \\ 
  \qquad WISDM Taste & 5 [3, 6] & 5 [3, 6] \\ 
  \qquad WISDM Tolerance & 5 [4, 6] & 5 [4, 6] \\ 
  \qquad WISDM Weight Control & 2 [1, 3] & 1 [1, 3] \\ 
  \qquad WISDM Affective Enhancement & 3 [2, 5] & 3 [2, 4] \\ 
  Baseline QSU-Brief, Factor 1 (QSU f1) & 16 [10, 25] & 16 [10, 24] \\ 
  Baseline QSU-Brief, Factor 2 (QSU f2) & 7 [5, 13] & 7 [5, 12] \\ 
  Baseline QSU-Brief, Total Score (Total QSU) & 24 [15, 37] & 24 [16, 35] \\ 
  MNWS Score & 5 [3, 10] & 5 [2, 9] \\ 
  PANAS Score (Positive) & 32 [25, 37] & 31 [26, 37] \\ 
  PANAS Score (Negative) & 14 [11, 19] & 13 [11, 17] \\ 
  CES Satisfaction & 5 [4, 6] & 5 [4, 6] \\ 
  CES Psychological Reward & 3 [2, 4] & 3 [2, 4] \\ 
  CES Aversion & 1 [1, 1] & 1 [1, 2] \\ 
  CES Enjoyment of Sensation & 3 [2, 5] & 4 [2, 4] \\ 
  CES Craving Reduction & 5 [4, 6] & 5 [3, 6] \\ 
\bottomrule
\end{tabular}
\caption{NNK, 4-(methylnitrosamino)-1-(3-pyridyl)-1-butanone, a tobacco-specific nitrosamine; PGEM, Prostaglandin E2 metabolite; FTND, Fagerström Test for Nicotine Dependence; CESD, Center for Epidemiologic Studies Depression; SMAST, Short Michigan Alcoholism Screening Test; DAST, Drug Abuse Screening Test; WISDM, Wisconsin Inventory of Smoking Dependence Motives; QSU-Brief, Brief Questionnaire of Smoking Urges; MNWS, Minnesota Tobacco Withdrawal Scale; PANAS, Positive and Negative Affect Schedule; CES, Cigarette Evaluation Scale. }
\label{demo}
\end{table}

\subsection{Context for Data Application }
Smoking remains the leading cause of preventable death in the US \cite{Warren, who2011, who2017}.
Through the Family Smoking Prevention and Tobacco Control Act, the Food and Drug Administration (FDA) has the authority to mandate a reduction in the nicotine content of cigarettes to non-addictive levels if it would benefit public health. Nicotine reduction is a promising method to facilitate quit attempts and consequently reduce cigarette-caused disease and death, a potential regulation that is currently being proposed and investigated. \cite{Hatsukami}.
Clinical trials have exhibited, on average, a significant difference between cigarettes containing 15.5mg/g tobacco (which is similar to what is observed in most cigarettes on the market) compared to 0.4mg/g tobacco in reducing smoking behavior among smokers. 
However, to fully understand the full public health impact of a nicotine reduction policy, it is important to investigate whether subgroups of participants within the trial may have differential treatment effects.

We considered data from 
a randomized, double-blind trial (n=1250) where adult smokers were assigned to 1 of 3 treatment conditions:
1. Immediate nicotine reduction, where participants were assigned to use cigarettes with a nicotine content of 0.4 mg/g tobacco. 
2. Gradual nicotine reduction, where participants smoked progressively lower nicotine content cigarettes for a period of one month each until receiving the same 0.4 mg/g tobacco VLNC cigarettes as the immediate reduction group for the last month of the study. 
3. Usual nicotine control, where participants were given their assigned cigarettes with a nicotine content of 15.5 mg/g tobacco, which is similar to what is observed in most cigarettes on the market.
Participants were given study cigarettes for a period of 20 weeks and outcome measures assessed included biomarkers of smoke exposure, and participants' measures of smoking behavior were recorded. 
The dataset contains 46 demographic and baseline covariates such as age, gender, race, education, cigarettes per day, total nicotine equivalents (TNE, a common measure of internal nicotine dose which includes nicotine and its metabolites), carbon monoxide levels, etc. 
Only the immediate nicotine reduction and the usual nicotine control groups were used in this analysis since the effect of VLNC cigarettes was much more pronounced in the immediate group than the gradual reduction group when compared to the control \cite{hatsukami2018effect}, and the average treatment effect in this case was $\sim$7 CPD.  
Table \ref{demo} provides summaries of baseline variables by treatment group; the two groups were well balanced for most variables, as would be expected after randomization. 
The outcome of interest in our analysis is the total cigarettes  
(study and non-study cigarettes) per day (CPD) smoked at week 20.

The goal of this analysis is to use the VT framework to predict individual treatment effects based on baseline covariates, and then segment the population into subgroups based on potential moderators of the effect of VLNC cigarettes.
The goal is not to identify subgroups of the population where the intervention of VLNC would cause participants to smoke more or less CPD than they would have if they were smoking control cigarettes by the end of the 20 week study; we are instead interested in describing the heterogeneity of the treatment effect in a parsimonious way. 

\subsection{Results}

\begin{table}[!htbp]
\caption{Predictors of VLNC treatment effect using various VT methods}
\centering
\begin{tabular}{m{2.3cm}cclc}
\toprule
& \textbf{LASSO} & \textbf{RF} & \textbf{MARS} & \textbf{Superlearner} \\
\midrule
\textbf{Regression Tree} &
\makecell[lc]{Baseline CPD\\ Age\\} & 
\makecell[lc]{Baseline CPD\\ Age\\} & 
\makecell[lc]{Age\\ Baseline Biomarker for NNK\\ Baseline CPD\\PANAS (Positive)\\WISDM Weight Control\\Baseline QSU f2\\PGEM\\} & 
\makecell[lc]{Age\\ Baseline CPD \\ Baseline QSU f1\\ Baseline TNE\\}\\
\midrule
\textbf{Conditional Inference Tree} &
\makecell[lc]{Baseline CPD\\ Age\\} & 
\makecell[lc]{Baseline CPD \\Age\\ PANAS (Negative)\\ Race\\} &
\makecell[lc]{Age\\ Baseline CPD \\Baseline QSU f2\qquad\qquad\qquad\quad\\} &
\makecell[lc]{Age\\ Baseline CPD \\Baseline TNE\\Baseline QSU f1\\}\\
\bottomrule
\end{tabular}
\label{cenic}
\end{table}

We applied the Step 1 VT methods discussed in Section 1 (LASSO, Random Forest, MARS and Superlearner) with the regression and conditional inference trees for Step 2 to evaluate TEH for this trial. 
The predictors that delineated the different subgroups from each of the VT analyses can be found in Table \ref{cenic}. The predictors selected from each Step 1 method are similar regardless of if regression trees or conditional inference trees were used for Step 2. Age was found to be an important variable in every method combination, as well as baseline CPD. Only the regression tree with individual treatment estimates from MARS identified more than four covariates to delineate the subgroups. 
An example regression tree with individual treatment estimates from Superlearner is shown in Figure \ref{extree}. Additional regression and conditional inference trees can be found in Appendix \ref{app_trees}. We can see that the first split occurs at age 47, with younger participants having a larger reduction in the estimated number of CPD. Further splits on the baseline CPD, baseline TNE, baseline QSU f1 and age divided the population into 8 subgroups, where QSU f1 is Factor 1 of the QSU-Brief, which is a measure of smoking craving. Factor 1 summarizes the craving to smoke with smoking perceived as rewarding, as compared to Factor 2, which summarizes the craving to smoke due to an anticipation of relief from negative affect \cite{qsubrief}. 
\begin{figure} [!htbp]
\centering
    \includegraphics[width=0.8\textwidth]{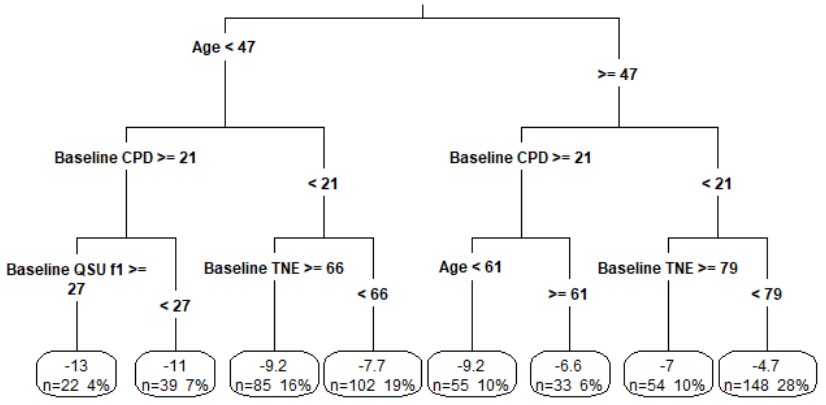}
    \caption{Example regression tree with individual treatment estimates from Superlearner}
    \label{extree}
\end{figure}

All the different method combinations gave from 4 to 8 subgroups. In all the trees constructed from VT using the various prediction methods, all of the subgroups had a treatment effect from a reduction of 1.7 CPD to a reduction of 13 CPD. There were no subgroups where participants were estimated to smoke more CPD than at baseline if VLNC cigarettes were used instead of the control cigarettes. Overall, participants under the age of 47 and smoking heavily at baseline have a more drastic decrease in the number of cigarettes smoked per day as compared to older participants who are smoking less at baseline.

\section{Discussion}
In both the linear and nonlinear cases, the performance of VT is heavily dependent on the modeling choice of step 1. 
In the linear case, VT implemented with the LASSO or Superlearner with the LASSO as a candidate model for step 1 had the best performance for all three metrics we considered. From the simulation results we obtained, using a linear model for step 2 of VT only outperformed the tree-based models slightly for the individual treatment benefit classification. 
In the nonlinear case, VT with Random Forest or Superlearner in step 1 yielded the best performance for all 3 metrics when paired with a regression tree for step 2. 
From these simulations, regression trees seems to have the most consistent performance in both the linear and nonlinear cases.
Interestingly, when LASSO was used for step 1, almost all the variables selected were truly predictive of the outcome for sample sizes of 1000 and 600.

The sample sizes also had an impact on the performance of VT. From our simulations, it can be seen that samples of 1000 and 600 gave largely accurate and consistent results if the ``correct" step 1 method, or a Superlearner with the ``correct" model included was used. 
However, in the 200 sample cases, the performance of all 3 metrics was inconsistent. In the linear case, using LASSO is still comparable to using a Superlearner model with LASSO as a candidate model for all 3 metrics we evaluated. However, in the nonlinear case, the Random Forest outperformed the Superlearner with Random Forest as a candidate model for all 3 metrics. 
For both the linear and nonlinear cases, the different problem settings of ``Regular", ``Correlated" and ``Selection Bias" gave similar results in most cases of our simulations. This indicates that VT is fairly robust to these types of data problems, and thorough pre-processing of the data is not necessary. 

However, our simulation study does have some limitations. 
The nonlinear data was generated using a tree structure. This biases the results towards the tree-based methods in the simulation.
Also, the correlated covariates and selection bias data scenarios we emulated in this paper do not usually exist in isolation, and there will usually be a combination of complicating factors. Our simulations break down the individual impact of each scenario, but a combination of these scenarios could cause VT to perform differently.  

The conclusions drawn from the simulation study are reflected in our analysis of the clinical trial dataset. Here again, the choice of method to use in Step 1 is more influential than the choice of Step 2; the regression tree and causal tree from the same Step 1 method are similar if not identical in most cases. Overall, the results of the VT seem generally invariant to the choices of Step 1 and 2 methods for the dataset. Though the tree constructed using the different method combinations are not identical, they all show a very similar story: while most participants are being placed in subgroups with an estimated treatment effect size of a decrease in CPD of ~7 compared to control, there are smaller subgroups of older participants with elevated baseline CPD or TNE that saw an even greater decrease in CPD of ~12. For younger participants who have a lower baseline CPD or TNE, most subgroups have an estimated decrease in the number of CPD of ~4. These results are consistent with our expectations; younger smokers have a shorter smoking duration and may be less dependent and more likely to change their behavior, while TEH due to the baseline rate of smoking is to be expected due to a ceiling on the treatment effect for light smokers at baseline.

In the future, it would be interesting to evaluate the nonlinear results of the simulation using different data generation schemes that were not tree based. In addition, since this study only examined the case where there is a binary treatment, it is unclear how these results would generalize to additional treatment groups.  

In summary, the choice of method for Step 1 has substantial impact on VT performance and, ideally, the method used in Step 1 could be matched to the anticipated features of the data. If no prior information is known, then a black-box model such as Superlearner should be considered since it performs just as well as the LASSO in the case where the data is linearly related to the outcome, and it performs adequately in nonlinear cases as well. Also, in cases where the sample size is less than 1000, special care should be taken with the interpretation of results. 

\section*{Data Availability Statement}

The R code to recreate the simulations in Section~\ref{sims} are available on GitHub. The data used in Section~\ref{application} are currently not publicly available. The Principal Investigator intends to make the data publicly available through the NIDA Data Share Website at a later time.

\bibliographystyle{ieeetr}
\bibliography{WileyNJD-AMA}

\begin{singlespace}
\newpage
\appendix
\addcontentsline{toc}{section}{Appendices}
\section*{Appendices}

\vspace{5cm}

\section{Data Generation} \label{app_DG}

\begin{table}[!htbp]
\centering
\tiny
\begin{tabular}{|m{2em}|c|c|c| } 
\hline
 &\makecell[c]{\\ \textbf{REGULAR}\\ \\} & \textbf{CORRELATED} & \makecell{\textbf{SELECTION}\\ \textbf{BIAS}} \\
\hline
\rotatebox{90}{\makecell{\textbf{CONTINUOUS} \\ \textbf{COVARIATES}}}
& 
\makecell[c]{
    For $i=1,...,n$ and $j=1,...,100$ \\
    \\
    \qquad $\begin{aligned}[t]
        \mu_{j} &\sim N \Big(0,  3\Big) \\
        \boldsymbol{X}_{i} &\sim N \Big(\boldsymbol{\mu}, \boldsymbol{I}_{100} \Big)\\
        & \text{where } \boldsymbol{\mu}=(\mu_1,...,\mu_{100})
    \end{aligned}$
}
& 
\makecell[lc]{
    For $i=1,...,n$ and $j=1,...,100$ \\
    \\
    \qquad $\begin{aligned}[t]
        \mu_{j} &\sim N \Big(0,  3 \Big) \\
        \boldsymbol{X}_{i} &\sim N \Bigg(\boldsymbol{\mu},  \begin{pmatrix}
        1 &0.7 &0.7 &0.7 &&\\
        0.7 &1 &0.7 &0.7 &&\\
        0.7 &0.7 &1 &0.7 && \text{\Large0} \\
        0.7 &0.7 &0.7 &1 &&\\
        &&&&&\\
        &\text{\Large0}&&&&  \boldsymbol{I}_{96}\\
        \end{pmatrix}\Bigg) \\
            & \text{where } \boldsymbol{\mu}=(\mu_1,...,\mu_{100})
    \end{aligned}$ 
}
&
 \makecell{\\ \\ \\ \\ \\ \\ Same as \\ Regular Case \\ \\ \\ \\ \\ \\ \\} \\
\hline
\rotatebox{90}{\makecell{\textbf{BINARY} \\ \textbf{COVARIATES}}}
& 
\begin{tabular}{@{}c@{}} \\ For $j=101,...,110$ \\ $\boldsymbol{C}_{i} \sim Bern \Big(0.7 \Big)$ \\ \\\end{tabular}
& 
Same as Regular Case & \makecell{\\ \\ \\ \\ Same as \\ Regular Case \\ \\ \\ \\} \\ 
\hline
\rotatebox{90}{\makecell{\textbf{TREATMENT} \\ \textbf{ASSIGNMENT}}}
& \makecell[c]{\\ $T_i \sim Bern \Big(0.5\Big) $ \\ \\} & Same as Regular Case & \makecell{\\ \\ \\ \\ Same as \\ Regular Case\\ \\ \\ \\ }\\ 
\hline
\rotatebox{90}{\makecell{\textbf{LINEAR} \\ \textbf{OUTCOME VARIABLE}}}
& 
\makecell[lc]{\\
$\begin{aligned}[t]
    Y_{0i} & \sim N\Big(\big[\boldsymbol{X}_i^T \ \boldsymbol{C}_{i}^T \big] \times \bm{\beta} ^T, 3\Big)\\
    & \text{where } \bm{\beta}=\begin{cases}
        1, & \text{for $\beta_2 \dots \beta_{16}, \beta_{101}, \beta_{102} $}\\
        0, & \text{otherwise} \end{cases} \\
    Y_{1i} & \sim N\Big(\big[\boldsymbol{X}_i^T \ \boldsymbol{C}_{i}^T \big] \times \bm{\beta} ^T, 3\Big)\\
    &\text{where } \bm{\beta}=\begin{cases}
        1, & \text{for $\beta_1 \dots \beta_{20}, \beta_{101} \dots \beta_{105} $}\\
        0, & \text{otherwise} \end{cases} \\
    Y_i & = Y_{0i} \times (1-T_i) + Y_{1i} \times T_i
\end{aligned}$ \\
\\
This results in a $R^2$ of 0.63 for $Y_{0i}$ and 0.7 for $Y_{1i}$.\\
\\
}
& 
\makecell[c]{
Same as Regular Case, \\ 
and this results in a $R^2$ of 0.65 \\
for $Y_{0i}$ and 0.71 for $Y_{1i}$.}
& 
\makecell{Same as \\ Regular Case}
\\
\hline

\end{tabular}
\end{table}

\begin{table}[!htbp]
\centering
\tiny
\begin{tabular}{|m{2em}|c|c|c| } 
\hline
 &\makecell[c]{\\ \textbf{REGULAR}\\ \\} & \textbf{CORRELATED} & \textbf{SELECTION BIAS} \\
\hline
\rotatebox{90}{\makecell{\textbf{LINEAR OUTCOME} \\ \textbf{VARIABLE NO TEH}}}
& 
\makecell[lc]{\\
$\begin{aligned}[t]
    Y_{0i} & \sim N\Big(\big[\boldsymbol{X}_i^T \ \boldsymbol{C}_{i}^T \big] \times \bm{\beta} ^T, 3\Big)\\
    Y_{1i} & \sim N\Big(2 + \big[\boldsymbol{X}_i^T \ \boldsymbol{C}_{i}^T \big] \times \bm{\beta} ^T, 3\Big)\\
    &\text{where } \bm{\beta}=\begin{cases}
        1, & \text{for $\beta_1 \dots \beta_{20}, \beta_{101} \dots \beta_{105} $}\\
        0, & \text{otherwise} \end{cases} \\
    Y_i & = Y_{0i} \times (1-T_i) + Y_{1i} \times T_i
\end{aligned}$ \\
\\
This results in a $R^2$ of 0.67 for $Y_{0i}$ and 0.7 for $Y_{1i}$.\\
\\
}
& 
\makecell[c]{
Same as Regular Case, \\ 
and this results in a $R^2$ of 0.63 \\
for $Y_{0i}$ and 0.7 for $Y_{1i}$.}
& 
\makecell{Same as \\ Regular Case}
\\
\hline
\rotatebox{90}{\makecell{\textbf{NONLINEAR} \\ \textbf{OUTCOME VARIABLE}}}
& 
\makecell[lc]{\\
$\begin{aligned}[t]
 Y_{0i} & \sim N(K, 1)\\
    & \text{where } K=\begin{cases}
        20, & \text{for $X_{i1}>\mu_1$ and $X_{i2}>\mu_2$}\\
        23, & \text{for $X_{i1}>\mu_1$ and $X_{i2}\leq\mu_2$} \\
        25, & \text{for $X_{i1}\leq\mu_1$ and $X_{i5}>\mu_5$} \\ 
        22, & \text{for $X_{i1}\leq\mu_1$ and $X_{i5}\leq\mu_5$} \\
        \end{cases} \\
 Y_{1i} & \sim N(K, 1)\\
    &\text{where } K=\begin{cases}
        22, & \text{for $X_{i1}>\mu_1$ and $X_{i2}>\mu_2$}\\
        20, & \text{for $X_{i1}>\mu_1$ and $X_{i2}\leq\mu_2$} \\
        25, & \text{for $X_{i1}\leq\mu_1$ and $X_{i5}>\mu_5$} \\ 
        23, & \text{for $X_{i1}\leq\mu_1$ and $X_{i5}\leq\mu_5$} \\
        \end{cases} \\
    Y_i & = Y_{0i} \times (1-T_i) + Y_{1i} \times T_i
\end{aligned}$ \\
\\
This results in a $R^2$ of 0.76 for both $Y_{0i}$ and $Y_{1i}$.
\\ \\}
& 
Same as Regular Case & \makecell{Same as \\ Regular Case}\\
\hline
\rotatebox{90}{\makecell{\textbf{NONLINEAR OUTCOME} \\ \textbf{VARIABLE NO TEH}}}
& 
\makecell[lc]{\\
$\begin{aligned}[t]
 Y_{0i} & \sim N(K, 1)\\
 Y_{1i} & \sim N(2 + K, 1)\\
    &\text{where } K=\begin{cases}
        22, & \text{for $X_{i1}>\mu_1$ and $X_{i5}>\mu_5$}\\
        20, & \text{for $X_{i1}>\mu_1$ and $X_{i5}\leq\mu_5$} \\
        25, & \text{for $X_{i1}\leq\mu_1$ and $X_{i6}>\mu_6$} \\ 
        23, & \text{for $X_{i1}\leq\mu_1$ and $X_{i6}\leq\mu_6$} \\
        \end{cases} \\
    Y_i & = Y_{0i} \times (1-T_i) + Y_{1i} \times T_i
\end{aligned}$ \\
\\
This results in a $R^2$ of ??? for both $Y_{0i}$ and $Y_{1i}$.
\\ \\}
& 
Same as Regular Case & 
\makecell{Same as \\ Regular Case}
\\
\hline

\rotatebox{90}{\makecell{\textbf{TRAINING} \\ \textbf{SET}}}
& Random Sampling & Random Sampling 
& 
\parbox{.2\textwidth}{
\begin{enumerate}[leftmargin=10pt]
    \item In the linear case, $ S_i=\sum_{j=15 \dots 18}X_{ij}$ and in the nonlinear case, $ S_i=\sum_{j=1,5}X_{ij}$
    \item Create two sets consisting 
    of the top half and bottom
    half of the population ranked 
    by $S_i$
\[H_B = {S_i \leq S_{(\frac{1}{2}n)} } \] 
\[H_T = {S_i > S_{(\frac{1}{2}n)} }\] 
    \item Take 25\% of the training set from a random sampling
    of $H_B$ and 75\% from a random sampling of $H_T$.
\end{enumerate}} 
\\
\hline
\rotatebox{90}{\makecell{\textbf{TESTING} \\ \textbf{SET}}} & \makecell[c]{\\ Random Sampling \\ \\}& Random Sampling & Random Sampling \\
\hline
\end{tabular}
\end{table}
\end{singlespace}

\newpage

\section{Simulation Results for N=80} \label{app_80}

\begin{table} [!htbp]
\centering
    \begin{subfigure}{0.45\textwidth}
    \centering
        \includegraphics[width=0.55\textwidth]{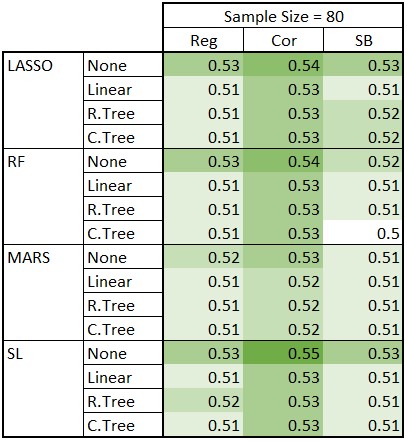}
        \caption{Linear case - Individual treatment benefit classification accuracy}
    \end{subfigure}
    \hspace{0.5cm}
    \begin{subfigure}{0.45\textwidth}
    \centering
        \includegraphics[width=0.55\textwidth]{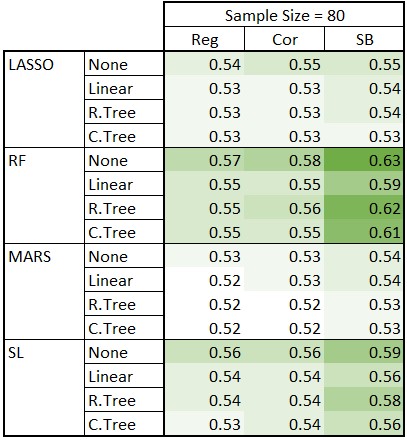}
        \caption{Nonlinear case - Individual treatment benefit classification accuracy}
    \end{subfigure} %
    
    \begin{subfigure}{0.45\textwidth}
    \centering
        \includegraphics[width=0.55\textwidth]{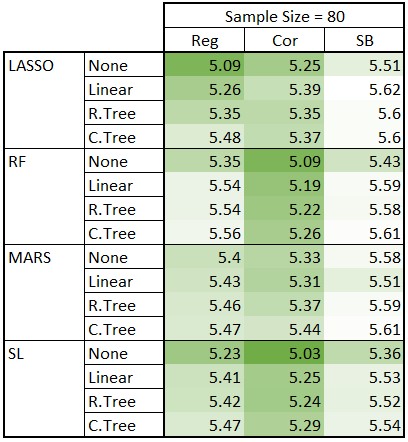}
        \caption{Linear case - Individual treatment effect MSE}
    \end{subfigure} %
    \hspace{0.5cm}
    \begin{subfigure}{0.45\textwidth}
    \centering
        \includegraphics[width=0.55\textwidth]{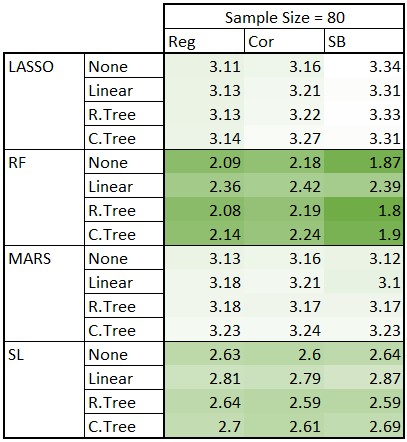}
        \caption{Nonlinear case - Individual treatment effect MSE}
    \end{subfigure}
    
    \begin{subfigure}{0.45\textwidth}
    \centering
        \includegraphics[width=0.55\textwidth]{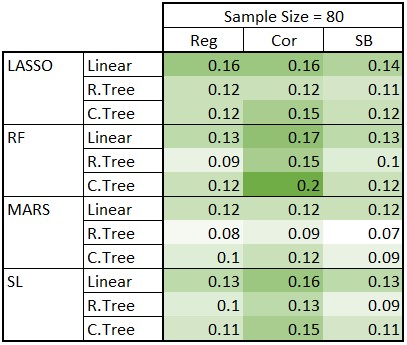}
        \caption{Linear case - Proportion of variables correctly selected by each model}
    \end{subfigure}
    \hspace{0.5cm}
    \begin{subfigure}{0.45\textwidth}
    \centering
        \includegraphics[width=0.55\textwidth]{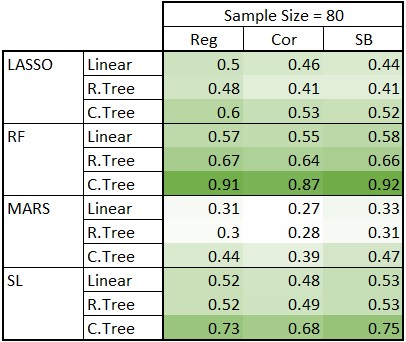}
        \caption{Nonlinear case - Proportion of variables correctly selected by each model}
    \end{subfigure}
    
\caption{Results of linear and nonlinear case simulations with a sample size of 80 under varying scenarios for 3 metrics. The rows indicate the methods used for step 1 and 2 of VT and the columns denote the different sample sizes and data settings. The labels of ``Reg", ``Cor" and ``SB" indicate the data generation of the regular case, correlated case and selection bias case described in section \ref{settings}. The labels of ``SL", ``R.Tree" and ``C.Tree" refer respectively to the Superlearner, regression trees and conditional inference trees used.}
\end{table}

\newpage
\section{Additional Trees from CENIC Data Application Analysis} \label{app_trees}

\begin{figure} [!ht]
\centering
    \begin{subfigure}{0.8\textwidth}
    \centering
        \includegraphics[width=0.8\textwidth]{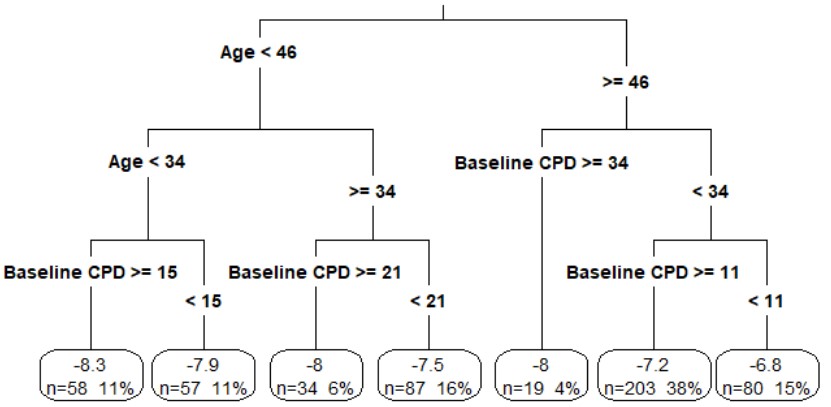}
        \caption{Regression tree with individual treatment estimates from LASSO}
    \end{subfigure}
    
    \begin{subfigure}{0.8\textwidth}
    \centering
        \includegraphics[width=0.8\textwidth]{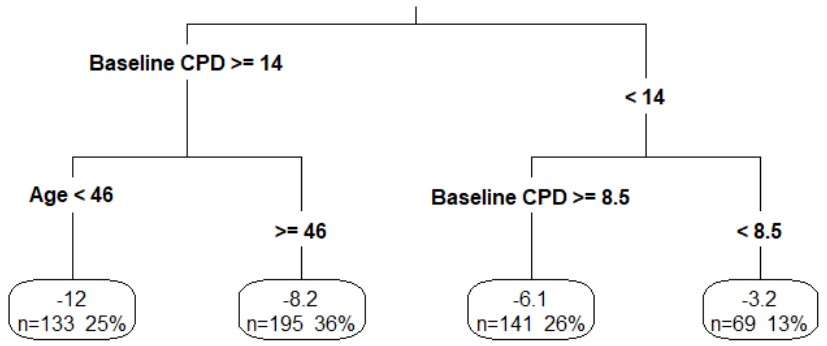}
        \caption{Regression tree with individual treatment estimates from Random Forest}
    \end{subfigure}
    
    \begin{subfigure}{0.8\textwidth}
    \centering
        \includegraphics[width=0.8\textwidth]{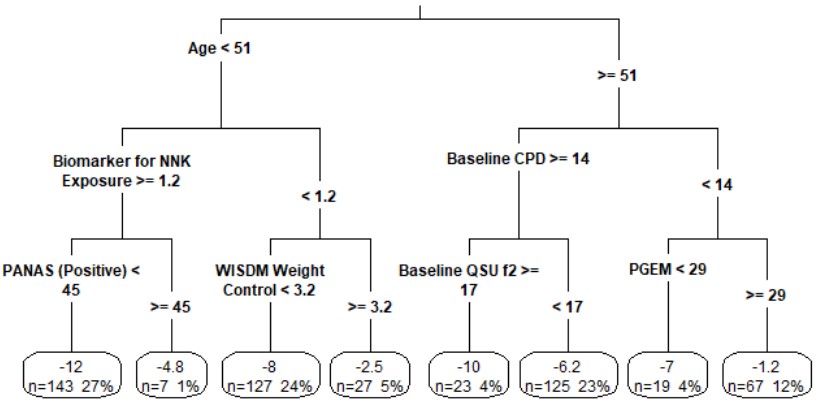}
        \caption{Regression tree with individual treatment estimates from MARS}
    \end{subfigure}

\caption{Resulting trees from VT by using various step 1 methods, and regression trees for step 2.}
\end{figure}

\begin{figure} [!ht]
\centering
    \begin{subfigure}{1\textwidth}
    \centering
        \includegraphics[width=0.9\textwidth]{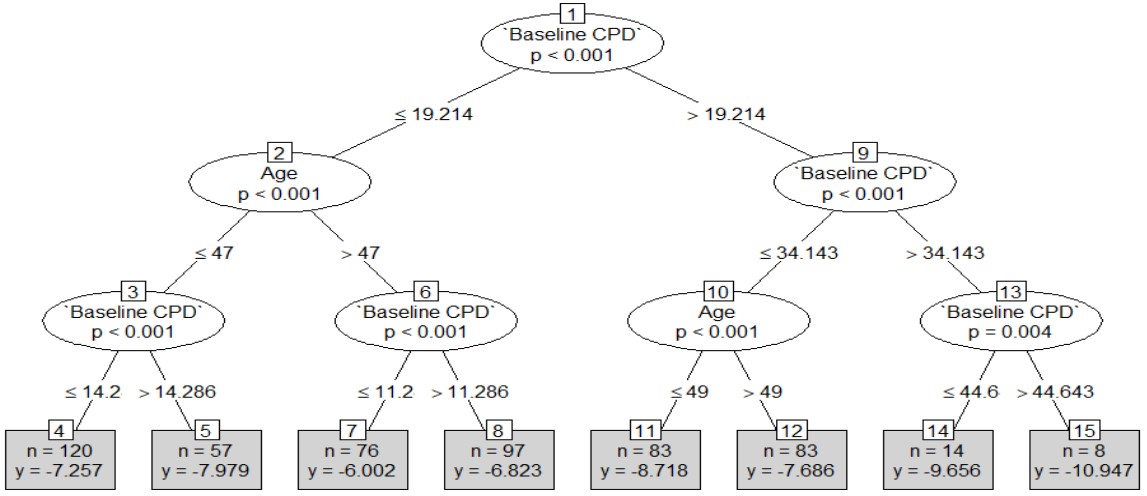}
        \caption{Conditional inference tree with individual treatment estimates from LASSO}
    \end{subfigure}
    \par\bigskip
    \begin{subfigure}{0.8\textwidth}
    \centering
        \includegraphics[width=0.8\textwidth]{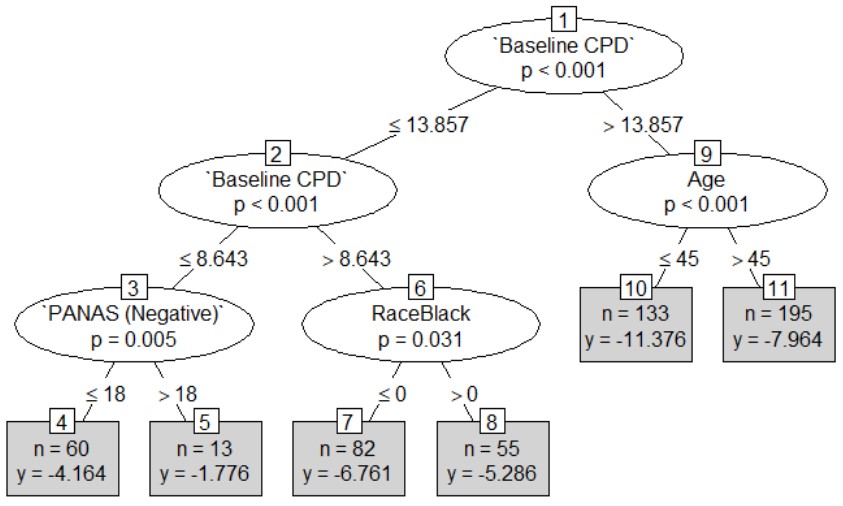}
        \caption{Conditional inference tree with individual treatment estimates from Random Forest}
    \end{subfigure}

\end{figure}

\begin{figure} [!ht] \ContinuedFloat
\centering
    \begin{subfigure}{1\textwidth}
    \centering
        \includegraphics[width=0.7\textwidth]{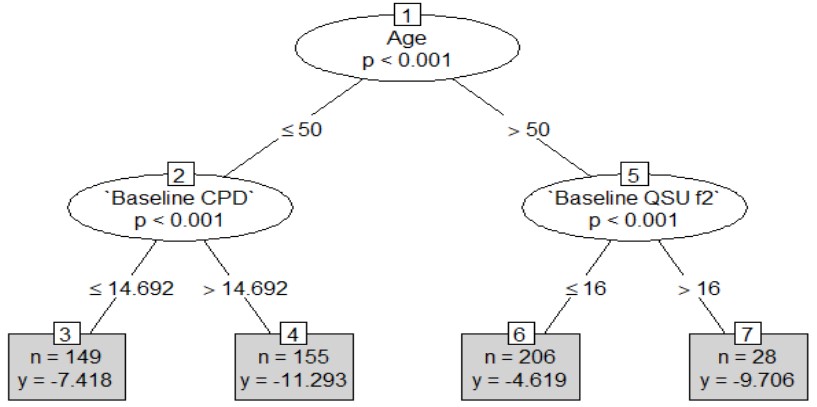}
        \caption{Conditional inference tree with individual treatment estimates from MARS}
    \end{subfigure}
    \par\bigskip
    \begin{subfigure}{1\textwidth}
    \centering
        \includegraphics[width=1\textwidth]{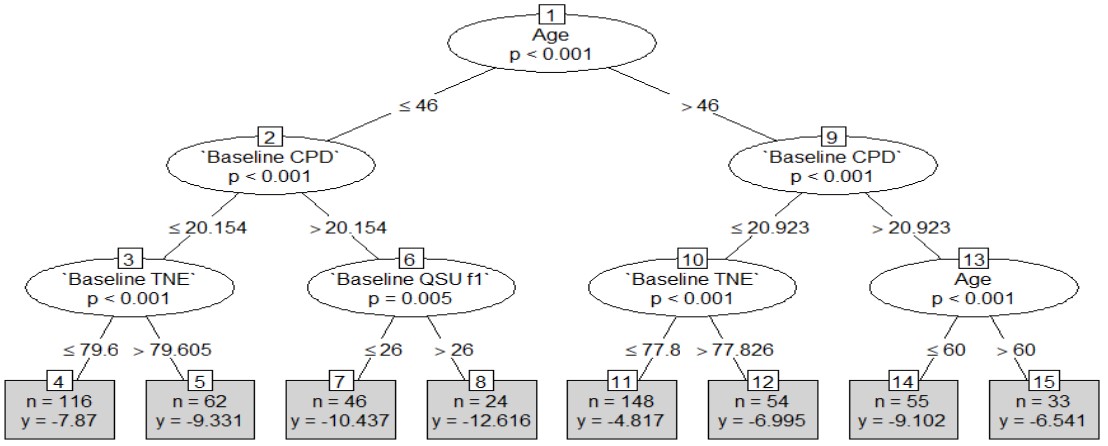}
        \caption{Conditional inference tree with individual treatment estimates from Superlearner}
    \end{subfigure}
    \caption{Resulting trees from VT by using various step 1 methods, and conditional inference trees for step 2.}
\end{figure}

\end{document}